# Neutron Diffraction Studies on Temperature Driven Crystallographic Anisotropy in $FeVO_4$ Multiferroic: Evidence of Strong Magnetostructural Correlations

Ajay Tiwari[1], Sagarmal Kumawat[1], Sudhindra Rayaprol[2] and Ambesh Dixit[1, a)]

[1]Department of Physics and C for Solar Energy, Indian Institute of Technology Jodhpur - 342037
[2]UGC-DAE Consortium For Scientific Research, Mumbai Centre, BARC Campus, Mumbai - 400085

[a)]Corresponding author: ambesh@iitj.ac.in

**Abstract.** We used temperature-dependent neutron diffraction measurements on $FeVO_4$ to understand the temperature driven anisotropy and observed that the maximum change for *a* and *b* lattice parameters in conjunction with a large contraction in angle *β* as a function of temperature. The least changes are observed for the *c* lattice parameter and in *γ* angle. From these structural parameters, it can be said that, $FeVO_4$ exhibits large structural anisotropy with lowering temperature. The large change in lattice parameters in magnetic phases i.e. below 22 K explains the strong magnetostructural coupling in $FeVO_4$.

## INTRODUCTION

$FeVO_4$ is a type II multiferroic system, showing the onset of ferroelectricity at ~ 15 K at the cost of the non-collinear spin structure of Fe atoms in triclinic crystallographic structure with $P\bar{1}$ symmetry. The low symmetry of $FeVO_4$ allows probing the crystallographic anisotropy, which may assist in understanding the onset of magnetic and ferroelectric phases. This system goes under several irreversible phase transitions under high pressure [1]. However, triclinic ($P\bar{1}$) structure is the most thermodynamically stable system under normal ambient conditions and does not exhibit any structural phase transition up to 3 GPa pressure [1]. This triclinic $FeVO_4$ undergoes two antiferromagnetic transitions at 22K and 15K from the high-temperature paramagnetic phase. The first antiferromagnetic transition at 22 K is incommensurate collinear, whereas the second transition at 15 K, followed by high temperature incommensurate collinear antiferromagnetic phase, is incommensurate non-collinear antiferromagnetic phase [2, 3]. More interestingly, the system also exhibits the onset of ferroelectric ordering at 15 K in conjunction with non-collinear antiferromagnetic phase simultaneously, making this system interesting [2, 3, 8, 9]. The detailed magnetic, dielectric and ferroelectric properties are investigated by Dixit et al [3], A. Daoud-Aladine et al [2], Kundys et al [10] and He et [11], and thermal conductivity by Dixit et al [4]. The system also exhibits strong magnetodielectric coupling not only in the multiferroic phase below 15 K but also exhibit at relatively higher temperatures [5]. This high-temperature magnetodielectric coupling is attributed to the probable magnon-phonon coupling, which is substantiated by the temperature dependent Raman studies [3,5].

Further, various non-magnetic and magnetic dopants are investigated at the iron site to understand their impact on magnetic and ferroelectric ordering. The system showed strong robustness against these dopants up to about 20% atomic fraction [6], persisting both antiferromagnetic ordering and low temperature (15 K) ferroelectric ordering. However, reduction in ferroelectric polarization is noticed with increasing the dopant atomic fraction at the iron site in $FeVO_4$. In spite of detailed ferroic properties and magnetodielectric properties, low-temperature structural properties are explored much. Aladine et al showed the onset of different magnetic ordering using temperature dependent neutron diffraction measurements and observing the magnetic Bragg-peak at 2 and 18 K. However, the variation of lattice parameters and associated asymmetry are not discussed in detail. We carried out temperature dependent neutron diffraction measurements from 300 K down to 3 K and strong temperature driven asymmetry is observed in crystallographic parameters.

## EXPERIMENTAL DETAILS

Bulk $FeVO_4$ sample is prepared using a two-step solid-state synthesis approach. Initially, the stoichiometric ratio of iron (III) oxide (red $Fe_2O_3$ powder from Merck >90% pure) and Vanadium (V) oxide ($V_2O_5$ ~99% pure from Alpha Aesar) are mixed homogeneously and heated at 350 $^0C$ for 3 hrs. The sample was ground and finally heated at 750 $^0C$ for 1 hr in the air at 2 $^0C$ heating rate with normal cooling. The brown color powder was collected for further experiments. The neutron diffraction measurements are carried out on PD-3 diffractometer at Dhruva Reactor, India using neutrons at wavelength, $\lambda = 1.48$ Å. Temperature dependent measurements are done using a closed cycler refrigerator (CCR) based cryostat. The diffraction data was indexed using LeBail method and refined Reitveld method using FULLPROF suite refinement program [7].

## RESULTS AND DISCUSSION

The room temperature (300 K) neutron diffraction data is shown in Fig 1 with refinement data. The refined structure is well matched with triclinic $P\bar{1}$ crystallographic structure substantiating the phase purity of bulk $FeVO_4$ sample. The estimated lattice parameters $a$, $b$ and $c$ are 6.73289 Å (±2.70966E-4Å), 8.06373 Å (±3.65915E-4Å), 9.31462 Å (±3.14378E-4Å), and angle $\alpha$, $\beta$ and $\gamma$ are 96.52226° (±0.00303°), 106.68535° (±0.00275°), 101.53638° (±0.00342°), respectively. These values are consistent with the earlier reports [1, 3].

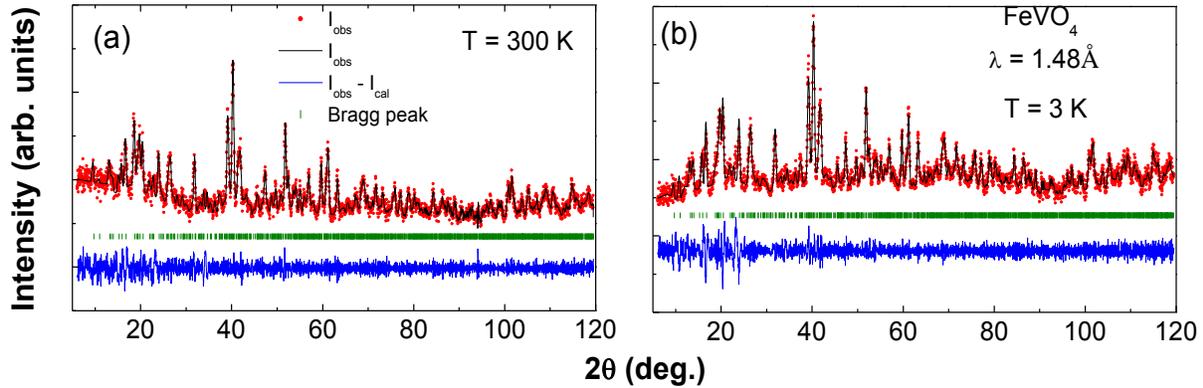

**FIGURE 1.** Neutron diffraction measurements and respective refinements at temperatures (a) 300 K and 3 K on bulk $FeVO_4$ powder sample

The neutron diffraction measurements are carried out at different temperatures using $\lambda = 1.48$ Å neutron beam and representative 3 K neutron diffraction data is shown in Fig 1 (b). The data is best fitted with triclinic $P\bar{1}$ crystallographic phase. The most surprising aspect in the present neutron diffraction measurement is the absence of Bragg peak corresponding to the magnetic structure. We attribute the noticed absence of magnetic Bragg peak due to the relatively large Q value with the present neutron wavelength used for the measurements. The long periodicity of magnetic correlations in $FeVO_4$ may require neutron diffraction measurements even at lower Q values, which may be achieved by using higher neutron wavelengths for diffraction experiments. In spite of being unable to nail down the magnetic structure of $FeVO_4$ below 22 K, we observed large variation in lattice parameters. The variation in lattice parameters is summarized in Fig 2. The variation in these parameters including the volume of the unit cell against temperature is shown in Fig 2 (left panel). A large variation in different lattice parameters is observed, which resulted in enhanced cell volume with lowering temperature, as can be seen in Fig 2. The right panel of Fig 2 is showing the zoomed view of low-temperature range 30 K – 3 K.

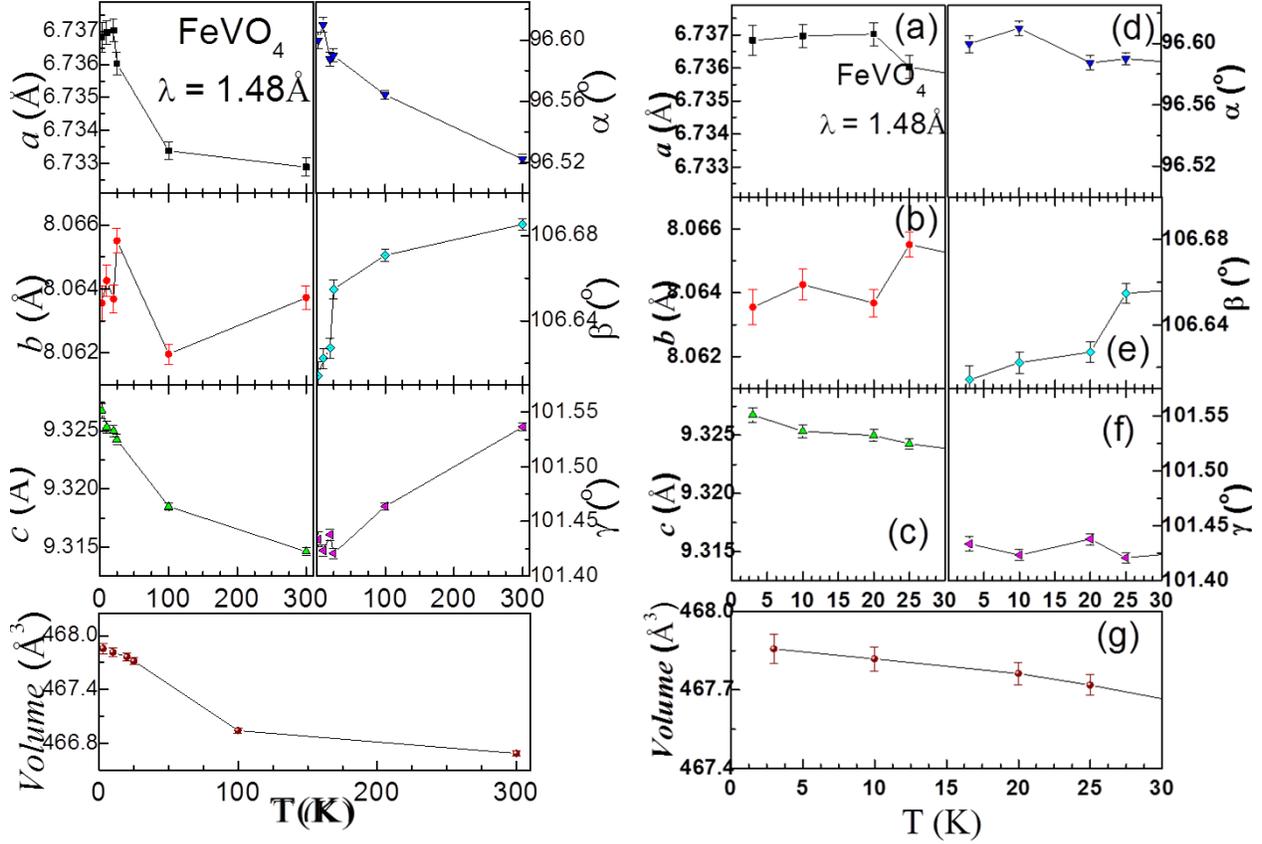

**FIGURE 2.** Variation in lattice parameters of bulk FeVO$_4$ against temperature: Left panel- 300 K – 3 K temperature range and the right panel showing the zoomed view in the low-temperature range 30 K – 3 K, with variation in volume against temperature

We observed that lattice parameter $c$ and angle $\gamma$ are less sensitive to the temperature, whereas other lattice parameters are showing strong variation against temperature in the low-temperature range. The lattice parameter $a$ and angle $\alpha$ are increasing initially with reducing temperature till 25 K and become nearly constant with any further reduction in temperature, Fig 2 (a & d) right panel. In contrast, the lattice parameter $b$ is the maximum near 25 K and reduced further with lowering temperature in conjunction with similar behavior for angle $\beta$, which is the maximum near 25 K and keeps on reducing with temperature. This variation in lattice parameters and lattice angles support the temperature driven crystallographic anisotropy in bulk FeVO$_4$, especially in the low-temperature region. The low-temperature region is also incommensurate antiferromagnetic and thus, this may be a signature of the onset of magnetostructural correlations in FeVO$_4$. The system is also ferroelectric at or below 15 K in lower non-collinear incommensurate antiferromagnetic phase, making system multiferroic, showing strong magnetodielectric coupling [2 - 5]. The microscopic origin of ferroelectric polarization requires the breaking of spatial inversion symmetry, and in the case of FeVO$_4$ non-collinear magnetic structure is supposed to be responsible for the same. This onset of ferroelectricity will lead to the structural changes in the lattice and the observed temperature-driven structural anisotropy may be the main source of the onset of ferroelectric ordering and strong magnetodielectric coupling in this system.

## CONCLUSION

We demonstrated temperature driven strong crystallographic anisotropy in bulk triclinic $P\bar{1}$ $FeVO_4$ using temperature dependent neutron diffraction. This substantiates the robust magnetostructural correlations, which may be the source of microscopic origin of ferroic ordering in conjunction with non-collinear incommensurate magnetic ordering below 15 K. However, we could not notice the onset of magnetic Bragg peak using $\lambda = 1.48$ Å neutron diffraction experiments due to higher Q limits. The experiments are in progress using $\lambda = 2.31$ Å, where lower Q range 0.28 - 4.71 ($Å^{-1}$) may show the onset of magnetic Bragg peak and thus, assist in understanding the detailed magnetic structure with other structural details simultaneously.

## ACKNOWLEDGMENTS

Author Ambesh Dixit highly acknowledges the financial assistance from UGC-DAE Consortium For Scientific Research, Gov. of India through project number CRS-M-221 for this work and Dr. V. Siruguri, Director UGC-DAE CSR, Mumabi for his encouragement and technical discussions.

## REFERENCES

1. S. L. Moreno et al, Inorg. Chem, 57, 7860-7876 (2018).
2. A. Daoud-Aladine *et al.* Physical Review B 80.22, 220402 (2009).
3. A. Dixit and G. Lawes. Journal of Physics: *Condensed Matter* 21.45, 456003 (2009).
4. A. Dixit et al. IEEE Transactions on Magnetics 51.11, 1-4 (2015).
5. A. Dixit, G. Lawes, and A. Brooks Harris. Physical Review B 82.2, 024430 (2010).
6. A. Kumarasiri *et al*. Physical Review B 91.1, 014420 (2015).
7. J. Rodrı́guez-Carvajal, Physica B 192, 55 (1993).
8. A. B. Harris, A. Aharony and O. E. Wohlman, J. Physics *Condensed Matter*, 20, 434202 (2008).
9. A. B. Harris Phys. Rev. B, 76, 054447-1-41 (2007).
10. B. Kundys, C. Martin, and C. Simon. Physical Review B 80.17, 172103 (2009).
11. Z. He and Jun-Ichi Yamaura et al, Journal of Solid State Chemistry 181.9, 2346-2349 (2008).